\documentclass[aps,prl,reprint,superscriptaddress,showpacs,floatfix,amsmath,amssymb]{revtex4-1}
\usepackage{graphicx}
\DeclareMathOperator{\trace}{Tr}
\newcommand{\be}{\begin{equation}}
\newcommand{\ee}{\end{equation}}
\newcommand{\nn}{\nonumber}
\newcommand{\rv}{\boldsymbol{ r }}
\newcommand{\rvp}{\rv^{\prime}}
\newcommand{\jv}{\boldsymbol{\jmath}}
\newcommand{\Ev}{\boldsymbol{E}}
\newcommand{\mat}[1]{ \underline{\boldsymbol #1} }
\newcommand{\ks}[0]{ \mathrm{s} }
\newcommand{\Hxc}[0]{ \mathrm{Hxc} }
\newcommand{\xc}[0]{ \mathrm{xc} }
\newcommand{\eq}[0]{ \mathrm{eq} }

\newcommand{\dyn}[0]{ \mathrm{dyn} }
\newcommand{\fder}[2]{ \frac{\delta#1}{\delta#2} }
\newcommand{\cE}[0]{ \mathcal{E} }
\newcommand{\intc}[0]{ \int_{\mathcal{C}} }
\newcommand{\ind}[1]{ \int\!\!\mathrm{d}#1 \; }

\begin{document}

\title{Density-Functional Theory of Thermoelectric Phenomena}

\author{F. G. Eich}
\email[]{eichf@missouri.edu}
\affiliation{Department of Physics, University of Missouri-Columbia, Columbia, Missouri 65211}

\author{M. \surname{Di Ventra}}
\affiliation{University of California, San Diego, La Jolla, CA 92093}

\author{G. Vignale}
\affiliation{Department of Physics, University of Missouri-Columbia, Columbia, Missouri 65211}

\date{\today}

\begin{abstract}
  We introduce a non-equilibrium density-functional theory of local temperature and associated local energy density
  that is suited for the study of thermoelectric phenomena. The theory rests on a local temperature
  field coupled to the energy-density operator. We identify the excess-energy density, in addition to the particle density, as the basic variable,
  which is reproduced by an effective noninteracting Kohn-Sham system. A novel Kohn-Sham equation emerges featuring a time-dependent
  and spatially varying mass which represents local temperature variations.
  The adiabatic contribution to the Kohn-Sham potentials is related to the entropy viewed as a
  functional of the particle and energy density. Dissipation can be taken into account by employing
  linear response theory and the thermoelectric transport coefficients of the electron gas.
\end{abstract}


\maketitle

\emph{Introduction --} Thermoelectric phenomena have long been the subject of intense research activity. More recently, renewed interest in these phenomena
has surfaced due to their implications in the development of sustainable energy sources \cite{NolasGoldsmid:01,DubiDiVentra:11}. Besides its practical importance, thermoelectricity raises a host
of fundamental questions and challenges. For instance, the thermopower of a given system is defined as the electric potential difference (at zero electrical current)
that is induced by a thermal gradient across it. In this case, the electronic system is in mechanical equilibrium, and yet there is a steady flow of heat.
At the microscopic level, we could argue that local temperature variations appear, which  must be related to the heat-current density, $\jv_q$. Unfortunately,
neither concept has an unambiguous microscopic definition \cite{DiVentra:08}. For the heat-current density the problem is that a unique local energy-density operator
does not exist \cite{ChettyMartin:92,LepriPoliti:03,WuSegal:09}.
Similarly,  the standard thermodynamic definition of temperature fails as soon as we leave the regime of local quasi-equilibrium (for an operational definition of local
temperature based on scanning thermal microscopy see, e.g., Refs.\ \onlinecite{DubiDiVentra:09,CasoLozano:10,BergfieldDiVentra:13}).
Achieving a clearer understanding of these quantities is important not only from the conceptual point
of view but also for the practical calculation of familiar quantities, such as the electrical resistance.  

In this Letter we propose a definition of the microscopic energy density and the associated temperature field, and we show that these quantities can be computed
through a theoretical scheme that directly generalizes the well-known
time-dependent density-functional theory (TDDFT) \cite{RungeGross:84,Ullrich:12}.
Our work is inspired by Luttinger's seminal paper on the thermoelectric transport coefficients of the homogeneous interacting electron gas \cite{Luttinger:64a}.
In the process of adapting the Kubo linear response formalism to thermal transport, Luttinger identified the ``gravitational field'' -- a field that couples linearly to the energy density --
as the mechanical proxy \cite{Shastry:09} of local temperature variations.
In particular, for small fields applied to an initially homogeneous electron liquid, one can write the linear response relations \cite{AshcroftMermin-13}, 
\be
\begin{pmatrix} - e \jv_{n} \\[1ex] \jv_{q} \end{pmatrix} =
\begin{pmatrix} L_{11} & L_{12} \\[1ex] L_{21} & L_{22} \end{pmatrix}
\begin{pmatrix} \frac{1}{e}\nabla \mu  - \nabla \phi \\[1ex] -\frac{\nabla T}{T} - \nabla \psi \end{pmatrix} ~, \label{e_th_transport}
\ee
where $-e$ is the charge of an electron, and ${\jv_{n}}$ and ${\jv_{q}}$ are the particle and the heat current, respectively.
The ${L_{ij}}$ are transport coefficients that describe the thermoelectric properties
of the system under investigation and they obey Onsager reciprocity relations, ${L_{ij}=L_{ji}}$.
From Eq.\ \eqref{e_th_transport} we can see that a charge current
is driven by a difference in chemical potential ${\mu}$ or an electric field ${\Ev = - \nabla \phi}$. Similarly, a heat current is induced by a gradient in temperature
or a gradient of Luttinger's ``gravitational field'', ${\psi}$. The fields $\phi$ and $\psi$ are the mechanical counterparts of the chemical potential and the temperature, respectively
\footnote{ Strictly speaking ${\psi}$ is the mechanical proxy for ${\delta T/(T+\delta T)}$. For a small temperature variations this implies ${\nabla \psi \sim T^{-1} \nabla T}$. }.
The virtue of including these fields in the Hamiltonian is that they enable us to obtain the transport coefficients from a microscopic calculation.

Luttinger's study was limited to linear response about a homogeneous liquid state. Here we develop Luttinger's idea into a full-fledged
density-functional theory (DFT) of inhomogeneous systems (such as, e.g., nanojunctions) that are driven out of equilibrium by time-dependent temperature fields and potentials.
Similar to ordinary DFT, we introduce the ``Kohn-Sham system'' -- a fictitious noninteracting system that reproduces the exact density and the exact excess-energy density (defined below)
caused by the varying temperature field.
The resulting Schr\"odinger-like equation for this system (Kohn-Sham equation) includes a spatially and temporally varying mass term of a form often encountered in theoretical studies
of compositionally graded semiconductors \cite{GellerKohn:93}.

We furthermore suggest two basic approximations for the Kohn-Sham potentials: the adiabatic local-density approximation and the linear response approximation -- the latter being
the simplest approximation that allows us to introduce dissipative effects. Last, we show how the linear response approximation
predicts thermal corrections to the electrical resistivity, in addition to the well-known viscosity corrections, which have attracted considerable attention in the recent
literature \cite{VignaleDiVentra:09,RoyDiVentra:11}.

\begin{figure}
 \begin{center}
   \includegraphics[width=3in]{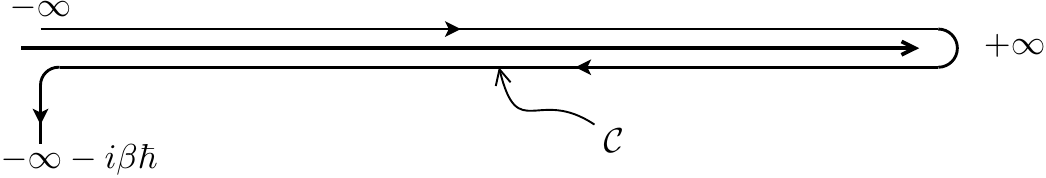}
 \end{center}
 \caption{Integration contour for the Keldysh action of Eq.\ \eqref{KeldyshAction}.}
 \label{KeldyshContour}
\end{figure}

\emph{Formulation --} One of the simplest ways to introduce thermal TDDFT is to start from the Keldysh action \cite{vanLeeuwen:98}, which we
slightly modify here to describe systems that evolve from an initial equilibrium state at temperature $T$.   The Keldysh action can be viewed
as the generalization of the thermodynamic potential, governing equilibrium
phenomena, to the time-dependent domain of non-equilibrium processes. Its form is
\be
A[v] \equiv i \hbar \ln\!\trace\!\left\{ T_\tau e^{- \frac{i}{\hbar} \intc \left[\hat  H+\ind{^3r} v(\rv,\tau)\hat n(\rv)\right]}\!\right\} ~, \label{KeldyshAction}
\ee
where the exponential is contour-ordered, as indicated by ${T_\tau}$ in Eq.\ \eqref{KeldyshAction},
along the path $\mathcal{C}=t(\tau)$ (parametrized by the real variable $\tau$, cf.\ Fig.\ \ref{KeldyshContour})
in the complex time plane and $\intc=\intc\!\mathrm{d}\tau t'(\tau)$.
In this formula $\hat H = \hat T+\hat W$ is the sum of kinetic energy, $\hat T$, and interaction energy, $\hat W$.
Further, $\beta =(k_BT)^{-1}$ (cf.\ Fig.\ \ref{KeldyshContour}) is the inverse temperature,
$\hat n(\rv)$ the particle density operator,  and $v(\rv,\tau) = V(\rv,\tau)-\mu$ the local time-dependent potential, $V(\rv,\tau)$, minus the chemical potential, $\mu$
\footnote{It is assumed that $v(\rv,t)$ tends to a constant $v_0(\rv)$ when $t \to -\infty$.  Then the definition \eqref{KeldyshAction} assumes that the system is initially
prepared in a thermal equilibrium state in the presence of the static potential  $v_0(\rv)$ at temperature $T$.}.
The time-dependent density is given by $n(\rv,\tau)=\delta A[v]/\delta v(\rv,\tau)$.
If the potential is time-independent, i.e. $v(\rv,\tau)=v_0(\rv)$ at all times, then the action functional, Eq.\ \eqref{KeldyshAction}, does indeed reduce to
$-i \beta \hbar \Omega[v_0]$, where $\Omega[v_0]$ is the grand-canonical thermodynamic potential.

We now want to extend the conventional TDDFT to allow for space- and time-dependent temperature fields.  The idea is to replace the global inverse temperature $\beta$,
which couples to the entire Hamiltonian, by a local temperature field $\beta[1 + \psi(\rv, t)]$, which couples to the local energy density $\hat h(\rv)$.
$\psi(\rv,t)$ (Luttinger's notation, cf.\ Ref.\ \onlinecite{Luttinger:64a}) is our ``temperature field''.  In the limit of slow spatial variation it naturally
describes the coupling of the system  to a local thermal reservoir at the corresponding temperature \cite{DubiDiVentra:09,CasoLozano:10,BergfieldDiVentra:13}.
It can thus be used for a first-principles treatment of electronic systems connected to local reservoirs at different temperatures and chemical potentials.
However, its microscopic significance is more general:
indeed, in a generic nonequilibrium situation one can define the  ``instantaneous local temperature'' in terms of the field $\psi$ that, when coupled to
the microscopic energy-density operator,  yields, under equilibrium conditions, the instantaneous local energy density of the nonequilibrium system.

The form of $\hat h(\rv)$ is not unique, since there are  different operators that,  integrated over $\rv$, produce the same Hamiltonian $\hat H$.
However, different forms are equivalent as far as their long-wavelength content is concerned. Here we choose
\begin{subequations} \label{energy_density}
  \begin{align}
    \hat h(\rv) & = \hat t(\rv)+\hat w(\rv)~, \label{h} \\
    \hat t(\rv) & = \frac{\hbar^2}{2m} \Big( \nabla_{\rv} \hat \phi^\dagger(\rv)\Big) \cdot \Big(\nabla_{\rv} \hat \phi(\rv)\Big) ~, \label{t} \\
    \hat w(\rv) & =\frac{1}{2} \ind{^3r'} \hat \phi^\dagger(\rv) \frac{\hat \phi^\dagger(\rvp) \hat \phi(\rvp)}{|\rv-\rvp|} \hat \phi(\rv) ~, \label{w}
  \end{align}
\end{subequations}
where $\hat t(\rv)$ and $\hat h(\rv)$ are the kinetic energy-density and interaction energy-density operator, respectively.

We are now ready to present our generalized  action functional,  which, extending Eq.\ \eqref{KeldyshAction},  reads
\begin{align}
& A[\tilde v,\psi] \equiv \nn \\
& i \hbar \ln\!\trace\!\left\{ \mathrm{T}_\tau e^{- \frac{i}{\hbar} \intc \left\{\hat H+ \ind{^3r} \left[\psi(\rv,\tau) \hat h(\rv)+\tilde v(\rv,\tau)\hat n(\rv)\right]\right\}}\!\right\} ~. \label{KeldyshAction2}
\end{align}
We have defined $\tilde v(\rv,\tau) \equiv  v(\rv,\tau)[1+\psi(\rv,\tau)]$: physically, this describes the coupling of the temperature field to the potential-energy density.
Equations \eqref{KeldyshAction} and \eqref{KeldyshAction2} highlight that ``density functionalization'' may be viewed as giving the intensive variables
$\mu \to v(\rv, t)$ and $\beta \to \psi(\rv, t)$ a space and time dependence. However, it is important to keep in mind that the corresponding densities, $n(\rv, t)$ and $h(\rv, t)$,
are, in general, not \emph{locally} related to $v(\rv, t)$ and $\psi(\rv, t)$.
The equations for the densities $(n,h)$ in terms of the potentials $(\tilde v,\psi)$ are
\be
n(\rv,\tau)=\fder{A[\tilde v,\psi]}{\tilde v(\rv,\tau)} ~,~~~ h(\rv,\tau)=\fder{A[\tilde v,\psi]}{\psi(\rv,\tau)} ~.
\ee
Inverting these equations yields (at least in the linear response regime \cite{Fernando:08})  a unique solution for the fields  $\tilde v(\rv,\tau)$ and $\psi(\rv,\tau)$
as functionals of  $n(\rv,\tau)$ and $h(\rv,\tau)$.
Legendre transformation of $A[\tilde v,\psi]$ with respect to $\tilde v$  and $\psi$ leads to the universal action functional $A[n,h]$
\footnote{$A[n,h]$ is defined as the \emph{negative} of the Legendre transform of $A[\tilde v,\psi]$.}.
The external potentials  $\tilde v$ and $\psi$ associated with the densities $n$ and $h$ are  given by the equations
\be \label{VariationalEquation}
\tilde v(\rv,\tau)= - \fder{A[n,h]}{n(\rv,\tau)} ~,~~~ \psi(\rv,\tau) = -\fder{A[n,h]}{h(\rv,\tau)} ~.
\ee

In order to make an explicit connection to Mermin's finite-temperature DFT (FT-DFT) \cite{Mermin:65} we now split $A[n,h]$ into two contributions:
an equilibrium part,  $A^\eq[n]$, which is easily related to the universal free-energy functional $F^\eq[n]$ of Mermin's equilibrium theory, and a remainder,
$\bar A[n,h]$, which we refer to as \emph{excess action}.  Thus, we write
\be  \label{A_Aeq_Abar}
A[n,h] =  A^\eq[n] + \bar A[n,h] ~,
\ee
where $A^\eq[n]=\intc F^\eq[n(\tau)]$.
Now, in view of the fact that  $F^\eq[n]=\ind{^3r}h^\eq[n](\rv) - \tfrac{1}{\beta}S^\eq[n]$, where $h^\eq[n](\rv)$ is
the equilibrium energy density and  $S^\eq[n]$ the equilibrium entropy for a given $n$,  we find it natural
to introduce a similar decomposition for the excess action, namely,
\be
\bar A[n,h] = \intc \ind{^3r}\bar h[n(\tau),h(\tau)](\rv) - \bar S[n,h] ~, \label{Abar}
\ee
where we introduced the \emph{excess}-energy density $\bar h[n(\tau),h(\tau)](\rv) \equiv h(\rv,\tau)-h^\eq[n(\tau)](\rv)$ (the energy density relative to the
\emph{instantaneous} equilibrium energy density) and the excess entropy functional
\be
\bar S[n,h] = S[n,h] - \frac{1}{\beta}\intc S^\eq[n(\tau)] ~, \label{Sbar}
\ee
where $S[n,h] \equiv \intc\ind{^3r}h(\rv,\tau) - A[n,h]$.
An important difference between the equilibrium entropy and the excess entropy is that the latter is nonlocal in time.
This means that it encodes retardation effects since it depends on the \emph{history} of the density and the energy density.

\emph{Kohn-Sham scheme --} A key concept in DFTs is the mapping of the interacting system onto a noninteracting system, the so-called Kohn-Sham (KS) system.
An important condition for the construction of the KS scheme in thermal DFT is that the usual KS scheme of Mermin's FT-DFT is reproduced
for equilibrium situations [$v(\rv, t)=v(\rv)$, $\psi(\rv,t)=0$]. In this way our theory is a true generalization of FT-DFT to nonequilibrium
situations described in terms of the charge- and energy-density variations.
The action functional $A_\ks[\tilde v_\ks,\psi_\ks]$ for the KS system is defined in complete analogy to Eq.\ \eqref{KeldyshAction2} by simply omitting
the contribution due to the electron-electron interaction in $\hat H$ \emph{and} $\hat h(\rv, t)$. Note that the operator $\hat h(\rv)$, yielding the
energy density of the interacting system, \emph{differs} from the operator
\be
\hat h_\ks(\rv) \equiv \hat t(\rv) = \frac{\hbar^2}{2m} \Big( \nabla_{\rv} \hat \phi^\dagger(\rv)\Big) \cdot \Big(\nabla_{\rv} \hat \phi(\rv)\Big)
\ee
representing the energy density of the KS system. In order to emphasizes this difference we will denote the energy density of the KS system
by $h_\ks$. Switching to the action functional $A_\ks[n,h_\ks]$ we obtain the KS potentials
\be
\tilde v_\ks(\rv,\tau)= -\frac{\delta A_\ks[n,h_\ks]}{\delta n(\rv,\tau)} ~,~~~\psi_\ks(\rv,\tau)=-\frac{\delta A_\ks[n,h_\ks]}{\delta h_\ks(\rv,\tau)} ~. \label{VariationalEquationKS}
\ee
Moreover $A_\ks[n,h_\ks]$ can be decomposed in the same way as $A[n,h]$ [cf.\ Eqs.\ \eqref{A_Aeq_Abar}--\eqref{Sbar}].
As in usual TDDFT the KS system reproduces the time-dependent density $n(\rv,t)$ of the interacting system.
The condition that the equilibrium limit of our theory coincides with FT-DFT is satisfied if the interacting and
the KS system have the same excess-energy density [cf.\ Eq.\ \eqref{Abar} and below].
Defining the Hartree-exchange-correlation ($\Hxc$) energy density
\be
\cE_{\Hxc}[n](\rv) \equiv h^\eq[n](\rv)-h^\eq_\ks[n](\rv)
\ee
we formulate our theory in terms of the KS energy density, by
viewing the interacting energy density as a functional of $n$ and $h_\ks$, i.e.,
\be
 h[n,h_\ks](\rv)=h_\ks(\rv)+\cE_\Hxc[n](\rv) ~.
\ee
Here we rely upon recent progress in constructing approximations to ${\cE_\Hxc[n]}$ derived from the uniform electron gas at arbitrary
temperatures \cite{PerrotDharma-wardana:00,BrownCeperley:13,BrownCeperley:13B,SjostromDufty:13,KarasievTrickey:13}.
The KS energy density is readily computed from the KS orbitals, which are obtained by solving
the time-dependent Schr\"odinger-like equation
\begin{widetext}
  \be
  i \hbar \partial_t  \phi_\alpha (\rv,t) = \left[- \hbar^2 \nabla_{\rv} \cdot \frac{1 + \psi(\rv,t) + \bar \psi_\xc(\rv,t)}{2m}  \nabla_{\rv}
    + \tilde v(\rv,t) + \tilde v^\eq_{\Hxc}(\rv,t) + \bar v_\xc(\rv,t) \right]\phi_\alpha(\rv,t) ~, \label{KohnShamEquation}
  \ee
\end{widetext}
starting from the noninteracting equilibrium state with inverse temperature $\beta$.
This is the KS equation for our theory.
Notice that the effective field $\psi + \bar \psi_\xc$ enters in this equation
as a position- and time-dependent correction to the effective mass.
In Eq.\ \eqref{KohnShamEquation} $\tilde v^\eq_{\Hxc}$ is the  equilibrium potential of Mermin's theory, $v^\eq_{\Hxc}$,  modified by the local temperature,
\begin{align}
\tilde v^\eq_{\Hxc}(\rv,t) =
v^\eq_{\Hxc}(\rv, t) +\! \ind{^3r^{\prime}} \psi(\rvp, t) \fder{\cE_{\Hxc}[n(t)](\rvp)}{n(\rv, t)}, \label{vHxc_eq}
\end{align}
and $\bar v_\xc$ and $\bar \psi_\xc$ are given as functional derivatives of the exchange-correlation ($\xc$) excess entropy, i.e.,
\begin{subequations} \label{bar_Potentials}
  \begin{align}
    & \bar v_\xc(\rv, t) = \!-\fder{\bar S_\xc[n,h_\ks]}{n(\rv, t)} , \bar \psi_\xc(\rv, t) = \!- \fder{\bar S_\xc[n,h_\ks]}{h_\ks(\rv, t)}, \label{vbar_psibar} \\
    & \bar S_\xc[n,h_\ks] = \bar S[n,h_\ks+\cE_{\mathrm{Hxc}}[n]] - \bar S_\ks[n,h_\ks] ~. \label{Sbar_xc}
  \end{align}
\end{subequations}
A detailed derivation of the KS potentials is provided in the Supplemental Material \footnote{The Supplemental Material can be found in
the source package provided at http://arxiv.org/format/1308.2311v3}.
From the time propagation of the KS orbitals we can readily compute the densities ${n = \sum_\alpha f_\alpha |\phi_\alpha|^2}$ and
the energy density ${h_\ks = \frac{\hbar^2}{2m} \sum_\alpha f_\alpha |\nabla_{\rv}\phi_\alpha|^2}$,
where $f_\alpha$ are the equilibrium occupations of the KS orbitals in the \emph{initial} ensemble.

We have already mentioned that the relation between the potentials $[ v(\rv, t), \psi(\rv, t) ]$
and the corresponding densities $[(n(\rv, t), h(\rv, t)]$ is \emph{nonlocal}.
Nevertheless, a large part of this nonlocality is accounted for by the solution of the noninteracting KS equation.
Indeed, the  success of DFT is largely due to the fact that the $\xc$ potentials can be approximated (semi-)locally
in space and time. Therefore we propose the following two approximations for our $\xc$ potentials:

\emph{Adiabatic local-density approximation --} The simplest approximation
-- the adiabatic local-density approximations (ALDA) -- ignores retardation and the remaining adiabatic contribution
is treated locally. Accordingly, the approximated potentials are \emph{functions} of the instantaneous densities, i.e.,
\begin{subequations} \label{ALDAPotentials}
  \begin{align}
    \bar v^{\mathrm{ALDA}}_\xc(\rv,t) & = - \frac{1}{\beta}\frac{\partial s_\xc(n(\rv, t), h_\ks(\rv, t))}{\partial  n(\rv, t)} ~, \label{vbar_xc_ALDA} \\
    \bar \psi^{\mathrm{ALDA}}_\xc(\rv,t) & = - \frac{1}{\beta}\frac{\partial s_\xc(n, h_\ks)}{\partial h_\ks(\rv, t)} . \label{psibar_xc_ALDA}
  \end{align}
\end{subequations}
In Eqs.\ \eqref{ALDAPotentials} $s_\xc(n, h_\ks)$ is the difference in entropy
of an interacting uniform electron gas with density $n$ and energy density $h_\ks + \cE_\xc(n)$ and a noninteracting uniform electron gas
with density $n$ and energy density $h_\ks$. Note that we also employ a \emph{local} approximation for $\cE_{\xc}[n]$ at finite temperature \cite{BrownCeperley:13B,KarasievTrickey:13}.
Furthermore, we show in the Supplemental Material that in a local approximation  $\tilde v^\eq_\Hxc \approx (1+\psi)v_\Hxc$. There we also present a more
detailed discussion of the adiabatic approximation.

\emph{Beyond the adiabatic approximation --}  The first step in going beyond the adiabatic approximation is to include the dependence
of the potentials on the time derivatives of the densities.  Since these time derivatives are related by continuity equations to
the divergence of particle and energy currents, it is natural to try and express the post-ALDA corrections in terms of current densities
\cite{Vignale:12,VignaleKohn:96-2,VignaleKohn:96,VignaleConti:97}.
The relevant currents are the particle current density $\jv_n=\frac{\hbar}{m}\Im m \sum_{\alpha} f_\alpha\phi^*_\alpha\nabla\phi_\alpha$
and the energy current density  $\jv_{h_\ks} = \frac{\hbar^3}{2m^2}\Im m \sum_{\alpha,i} f_\alpha (\partial_i \phi^*_\alpha)\nabla (\partial_i \phi_\alpha)$,
determining the full heat current density $\jv_q \equiv  \jv_{h_\ks} + (v^\eq_0 + v^\eq_\Hxc) \jv_n$
\footnote{These expressions neglect certain nonlinear terms involving the $\psi$ field and are therefore only appropriate in the linear response regime.}.
In the spirit of the local density approximation we use the thermoelectric conductivity matrix $\mat{L}$ of the homogeneous electron gas
(and its noninteracting version $\mat{L}_\ks$) to relate the gradients of the dynamical $\xc$ potentials
(i.e., the $\xc$ electric and thermal gradient fields) to the  particle and heat currents. Employing Eq.\ \eqref{e_th_transport} we thus have ($e^2=1$):
\be
\begin{pmatrix} \nabla \bar v^\dyn_\xc(\rv, t) \\[1ex] \nabla \bar \psi^\dyn_\xc(\rv, t) \end{pmatrix} =
\Big( \mat{L}_\ks^{-1} - \mat{L}^{-1} \Big) \begin{pmatrix} \jv_n(\rv, t) \\[1ex] \jv_q(\rv, t) \end{pmatrix} ~. \label{DynamicalXC2}
\ee
The structure of the thermoelectric resistivity matrix $\mat{L}^{-1}$ for the homogeneous electron gas is well known \cite{AshcroftMermin-13}:
\be
\mat{L}^{-1} = - \begin{pmatrix} \rho + \frac{\beta \Pi^2}{\kappa} & \frac{\beta \Pi}{\kappa} \\[1ex] \frac{\beta \Pi}{\kappa} & \frac{\beta}{\kappa} \end{pmatrix} ~, \label{resistivity}
\ee
where $\rho$ is the electrical resistivity, $\kappa^{-1}$ the thermal resistivity, and ${\Pi}$ the Peltier coefficient.
In a clean noninteracting electron gas both $\rho$ and $\kappa^{-1}$ vanish, reflecting the absence of scattering mechanisms.
In the interacting electron gas, however, the situation is profoundly different: on the one hand,  $\kappa^{-1}$ acquires a non-zero value, reflecting the
intrinsic decay of thermal currents caused by electron-electron interactions;  on the other hand, the homogeneous resistivity $\rho$ is replaced by viscous
friction  via the substitution $\rho \to - n^{-1}\nabla \eta \nabla n^{-1}$, where $\eta$ is the electronic viscosity due to electron-electron interactions \cite{VignaleDiVentra:09}.
The final formula for the non-adiabatic potentials is
\be\label{DynamicalXC3}
\begin{pmatrix} \nabla\bar v_\xc^\dyn \\[1ex] \nabla \bar \psi_\xc^\dyn \end{pmatrix} = \begin{pmatrix}
  -\frac{1}{n}\nabla \eta \nabla\frac{1}{n} + \frac{\beta{\Pi}^2}{\kappa} & \frac{\beta \Pi}{\kappa} \\[1ex]
  \frac{\beta \Pi}{\kappa} & \frac{\beta}{\kappa} \end{pmatrix}
\begin{pmatrix} \jv_n \\[1ex] \jv_q \end{pmatrix} ~,
\ee
where we have omitted the arguments $(\rv,t)$ for brevity. These potentials must be \emph{added} to the ALDA potential to constitute the full $\xc$ potentials
\footnote{Notice that Eq.\ \eqref{DynamicalXC3} agrees with the Vignale-Kohn formula (cf.\ Ref.\ \onlinecite{VignaleKohn:96,VignaleConti:97}) for the dynamical
$\xc$ potential for the ordinary (longitudinal) current channel.}.

\emph{Thermal corrections to the dc resistivity --}  Post-ALDA corrections allow us to study dissipative effects, such as the thermal contributions to the dc resistance of a conductor.
If $I$ is the current, then the energy dissipated per unit time is $W=RI^2$ where $R$ is the resistance of the conductor.  To calculate $R$ we perform a microscopic
calculation of the dissipated power. This is the work done by the external fields on the currents.
The dissipated power can be decomposed into a KS part and an $\xc$ correction \cite{VignaleDiVentra:09}, where
the former part is well described by the Landauer-B{\"u}ttiker formalism \cite{Landauer:57,BuettikerPinhas:85,Landauer:89}.
The $\xc$ correction to the dissipation reads,
\be
W_\xc = \ind{^3r} \left\langle  \jv_n(\rv) \cdot\nabla \bar v_\xc(\rv) + \jv_q(\rv)\cdot\nabla \bar \psi_\xc(\rv)  \right\rangle ~,
\ee
where the angular brackets denote the time-average over a period of oscillation of the fields, which tends to infinity ($\omega=0$) at the end of the calculation.
Only the part of the effective field that oscillates \emph{in phase} with the current, contributes to dissipation.
The effect comes entirely from the dynamical contribution to the $\xc$ fields, $\nabla \bar v_\xc^\dyn$ and
$\nabla \bar\psi_\xc^\dyn$, defined in Eq.\ \eqref{DynamicalXC3}, i.e.,
\be
W_\xc =  \ind{^3r} \left\langle  \eta \left| \nabla\frac{\jv_n}{n}(\rv) \right|^2 + \frac{\beta}{\kappa} \left|\jv_q(\rv) + \Pi \jv_n(\rv) \right|^2  \right\rangle ~.
\ee
More precisely, the first term corresponds to the so-called ``viscosity correction'' (cf.\ Ref.\ \onlinecite{VignaleDiVentra:09})
and the second term creates a ``thermal  correction'' to the Landauer-B\"uttiker result.

\emph{Conclusion --} The main accomplishment of this paper is formal: we have proposed a time-dependent density functional formalism for the study of thermoelectric phenomena and
we have  suggested two basic approximation strategies -- the adiabatic LDA and the linear response formalism.
We emphasize that the proposed thermal DFT focuses on the electrons. Phonons have not been included so far. We expect this to be adequate for nanoscale systems
where the electron-electron interactions dominate over electron-phonon or electron-impurity scattering \cite{AndreevSpivak:11}. For weak electron-phonon couplings
the phononic contribution to the heat current can be added to the electronic contribution perturbatively \cite{DiVentra:08}.
The effect of phonons on the electronic contribution may be taken in to account
by referring the local approximations proposed in this Letter to a uniform electron gas coupled to phonons.  Corrections due to impurities
have already been included in a similar manner \cite{UllrichVignale:02}.
We believe that our theory will enable the inclusion of electron-electron interaction effects in thermal transport with relative
ease when the required inputs, i.e., the entropy and transport coefficients for the homogeneous electron gas, become available.

\begin{acknowledgments}
  \emph{Acknowledgments --}  We gratefully acknowledge support from DOE under Grants No. DE-FG02-05ER46203 (F. G. E., G. V.) and DE-FG02-05ER46204 (M. D.).
\end{acknowledgments}

\bibliography{ThermalDFT-PRL}

\end{document}